# Enhancing 6G Network Security and Incident Response through Integrated VNF and SDN Technologies

Abdul Razaque[a], Abitkhanova Zhadyra Abitkhanovna[b]

[a]Department of Computer Science and Information Sciences, Arkansas Tech University, USA
[b]Department of Cybersecurity, Information Processing and Storage, Satbayev University, Kazakhstan

**Abstract**: Low-speed internet can negatively affect incident response in several ways, including decreased teamwork, delayed detection, inefficient action, and elevated risk. Delayed data acquisition and processing may result from inadequate internet connectivity, hindering security teams' ability to obtain the necessary information for timely and effective responses. Each of these factors may augment the organization's susceptibility to security incidents and their subsequent ramifications. This article establishes a virtual network function service delivery network (VNFSDN) through the integration of virtual network function (VNF) and software-defined networking (SDN) technologies. The VNFSDN approach enhances network security effectiveness and efficiency while reducing the danger of breaches. This method assists security services in rapidly assessing vast quantities of data generated by 6G networks. VNFSDN adapts dynamically to changing safety requirements and connection conditions through the use of SDN and VNF. This flexibility enables enterprises to mitigate or halt the impact of cyberattacks by swiftly identifying and addressing security threats. The VNFSDN enhances network resilience, allowing operators to proactively mitigate possible security attacks and minimize downtime. The incorporation of machine learning and artificial intelligence into VNFSDN can significantly improve network security and threat detection capabilities. The VNFSDN integrates VNF and SDN technologies to deliver security services that analyze vast quantities of 6G data in real time. As security requirements and network conditions evolve, it adapts dynamically to enhance network resilience and facilitate proactive threat detection. The VNFSDN further improves the scalability and adaptability of security services in 6G networks. Ultimately, a comparison with various VNFs, including Open Daylight, ONOS, and Cisco ACI, demonstrates that VNFSDN yields superior outcomes.

## 1. Introduction

Inadequate Internet connectivity during an incident response can result from various factors, including outdated equipment, network congestion, and an excessive number of connected devices [1]. This could lead to unstable networks, sluggish upload and download rates, and excessive latency, all of which could have a negative influence on incident response [2]. A faulty internet connection can cause a variety of problems, including dissatisfied clients, missed opportunities, slow response times, and lower productivity [3]. There are several solutions to the problem of internet speed in incident response. Software-defined networking (SDN) is one approach to increasing network reliability, adaptability, and efficiency [4]. SDN enables centralized network. Furthermore, replacing software and hardware and optimizing network arrangements are efficient approaches to improve network performance [5-6]. Another possible method for increasing internet speed is the use of virtual network functions (VNFs), which may be installed on virtual machines to optimize network architecture and improve overall network performance [7]. VNFs help enterprises effectively grow and manage their network resources, distribute services more effectively, and boost network performance [8]. Fast internet is essential for responding to incidents; yet a slow internet can result in missed chances, unsatisfied clients, and lost productivity [9]. Many devices linked to a network, network congestion, and antiquated hardware are the reasons behind poor internet connections [10]. Reducing slow internet speeds requires workable solutions in order to improve incident response times. The convergence of VNF and SDN technologies can result in a scalable, adaptable network infrastructure that is simple to manage and orchestrate, as shown in Figure 1. The two primary parts of the approach's structure are VNF and SDN, as seen in the figure. Switches and routers are examples of intermediary devices that are deployed in the cloud using the VNF. The SDN's primary purpose is to centrally manage intermediary devices. 6G is used throughout the project's construction, which expedites every step.

To address the issue of sluggish internet speeds, the Python platform is integrated with VNF SDN technology for a faster reaction time. With Python and VNFSDN, new applications for network security may be developed and evaluated. The suggested solution offers a possible approach to improving the network's overall performance and solving the issue of slow internet speed. It might also make centralized administration and enhanced security possible. The following is a summary of the study's primary contributions:

- The proposed VNFSDN investigates the application of SDN to enhance network scalability, flexibility, and efficiency. Other efficient methods to improve network performance include upgrading hardware and software and refining network configurations.
- VNFs optimize network architecture and improve network performance. To improve network security management and response times, the suggested VNFSDN combines VNFs with SDN features. Improved network security, cost savings, and centralized control are just a few advantages of this system. A versatile and adaptable method for examining network traffic and identifying security risks is offered by the Python platform used to create VNFs.
- There are two suggested algorithms. Utilizing VNFs and SDN technology, the first algorithm is a packet filtration utilizing VNFSDN that is compatible with the proposed 6G network security. It is suggested to use the second algorithm, packet capture and saving using the VNFSDN algorithm, to set up the variables and configurations required for packet capture and saving.

The remainder of the article is structured as follows:
Section 2 provides a summary of related work and reviews previous research on improving network performance and incident response times. Section 3 highlights the proposed plan. The proposed solution combines SDN and VNF to achieve a faster incident response and improved security. Section 4 discusses the experimental results and implementation,` and the results of testing the proposed solution. Section 5 discusses the results and an analysis and interpretation of the experimental results. Section 6 concludes the paper by summarizing the main findings and contributions of the study and recommends suggestions for future research.

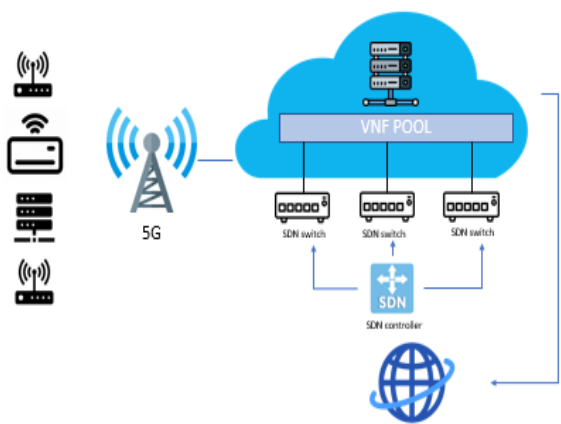

**Figure 1**: *General concept of software-defined networking with 6G*

*1.1. Problem Identification and Significance*

Insufficient internet bandwidth can cause lags in incident response times, which can prolong security breaches and cause more serious harm to an organization's money, reputation, and intellectual property. Slow internet speeds are therefore a major issue for companies and organizations, and it is imperative that workable solutions be developed to deal with this [14]. The problem of sluggish internet connections can be solved in a number of ways, such as by implementing SDN and VNFs, improving network configurations, and updating hardware and software [15]. Furthermore, to improve network performance and incident-response capabilities, current research has investigated the use of machine-learning techniques, edge computing, quality of service (QoS), and hybrid cloud architectures [16]. The response times of businesses and organizations to incidents can be greatly impacted by slow internet speeds, which can have fatal repercussions. Finding workable solutions to this issue is therefore essential. SDN and VNF technologies can be combined with machine learning techniques, edge computing, QoS implementation, and hybrid cloud architectures to provide strong and adaptable solutions for enhancing network performance and incident response capabilities.

## 2. Literature Survey

This section discusses the salient features of existing approaches. Abubakar and Pranggono [17] proposed machine learning as a solution for SDN-based intrusion detection and prevention. They further explored and highlighted the benefits and challenges of the proposed approach. They concluded that although machine learning can improve the accuracy of intrusion detection and reduce false positives, challenges related to scalability and training data availability still exist. Ahmed et al. [18] proposed VNF chaining and network slicing as possible solutions. The authors also mention their respective benefits and limitations. Research on the VNF sphere was introduced by Wang and Zhao [19] who explored the use of edge computing to improve network performance and address the challenges of latency and bandwidth requirements in incident responses. The authors outlined the architectures, use cases, and difficulties of edge computing while emphasizing how these technologies could speed up response times to incidents and ease network congestion. The authors outlined the architectures, use-cases, and difficulties of edge computing while emphasizing how these technologies could speed up response times to incidents and ease network congestion. Contributing to the development of QoS in SDN networks, Karakus and Durresi [20] noted that it might enhance reaction times and network performance. They investigated different QoS strategies and how well they worked to reduce network congestion and enhance QoS in SDN. Blockchain-based collaborative software-defined networking (BlockCSDN) was the subject of another pertinent study conducted by Li et al. [21]. The authors suggested enhancing network security and lowering the danger of cyberattacks by implementing blockchain technology into SDN. They talked about how blockchain may improve incident response skills and offer a tamper-proof record of network activity. In a study on network function virtualization (NFV) by Al-Najjar et al. [22], the authors proposed NFV to improve network efficiency, flexibility, and scalability and discussed the challenges related to resource allocation, network management, and security. The authors provided valuable insights into the current state of NFV and their potential for future networking technologies. Xu et al. [23] proposed hybrid cloud computing as a method of improving network performance. This study addressed the difficulties with security, privacy, and interoperability while concentrating on the possible cost- and scalability-saving advantages of hybrid cloud computing. Research in the field of VNF by Basu et al. [24] tackled the issue of constrained network storage and capacity, which can impair network QoS. The authors suggested a dynamic VNF sharing strategy known as FlexShare-VNF in order to maximize the placement of VNF instances over service function chains (SFCs) for better service delivery. In order to assign VNFs more effectively, Kim and Kim's research used a VNF placement approach that utilized VNF characteristics and information about each node's resources [25]. The authors proposed a way to assign VNFs promptly upon request and to find the right node for placement ahead of time by using periodic data on resource updates before VNF placement. Taniguchi and Shinomiya [26] proposed virtualized networks to minimize computing and network resources in the event of VNF failures. By reducing the cost of processing and network resources, the suggested approach seeks to assure sustainability against repeated VNF failures, which have the potential to seriously harm the network. 6G network speed and efficiency can be greatly increased by integrating VNFs with SDN technologies. VNFs can be dynamically deployed and operated to suit the unique requirements of various network functions and services by utilizing the programmability and flexibility of SDN. This method enhances the distribution of resources, reduces network congestion, and enhances security by enabling the implementation of advanced network policies and protocols. Hussain and Ali proposed a new approach using intrusion detection systems (IDS) [27]. They proposed a combined approach for IDS that leveraged both signature-based and anomaly-based techniques to enhance network security. The authors proposed that integrating both approaches would result in an IDS that is more effective and efficient because they understood the drawbacks of utilizing just one. To identify unknown attacks, their suggested method combined an anomaly-based technique with a signature-based approach for known attack detection. Raja [28] proposed a network firewall for cybersecurity in an IoT environment. The authors proposed that integrating both approaches would result in an IDS that is more effective and efficient because they understood the drawbacks of utilizing just one. To identify unknown attacks, their suggested method combined an anomaly-based technique with a signature-based approach for known attack detection.

## 3. Proposed 6G Security with VNF and SDN Technology

The suggested method improves network security by fusing Python and SDN's benefits. Network managers can install virtualized security features like firewalls and intrusion detection systems and centrally administer the network thanks to SDN. Python offers a versatile and potent framework for creating VNFs that analyze network data and identify security risks. The suggested strategy offers a scalable and effective way to manage network security by utilizing Python and SDN. There are various benefits to the previously mentioned VNF deployment algorithm. This enables security functions to be deployed and managed by network administrators as virtualized services. Moreover, expenses are decreased, and network security effectiveness is raised. A versatile and adaptable method for examining network traffic and identifying security risks is offered by the Python platform used to create VNFs. Network administrators may manage network resources from a single point of control thanks to SDN's centralized administration. As a result, network security is increased overall, and complexity is decreased. All things considered, the suggested algorithm offers a strong and effective method of network security that is ideal for meeting the needs of contemporary enterprises and organizations.

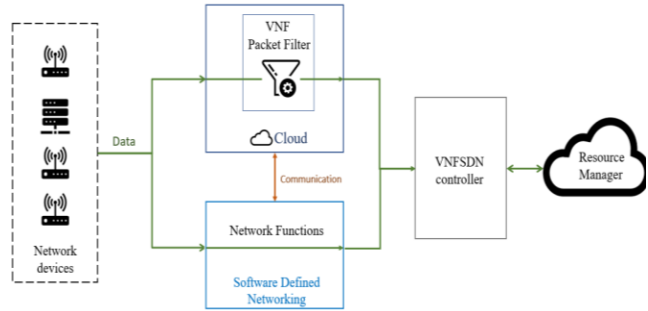

**Figure 2:** *The schema of the proposed VNFSDN solution*

Figure 2 shows the SDN controller as a VNF responsible for controlling the entire network infrastructure. The controller is connected to a centralized orchestration layer that manages the allocation of computing resources to different network functions. The VNFSDN approach consists of four main phases:
- Defining the VNF and SDN components and variables.
- Establishing communication between VNF and SDN.
- Decisions for operations and monitoring.

The VNFSDN approach is described in Algorithm 1.

**Algorithm 1:** Packet filtration using VNFSDN
**Initialization:** {V: Virtual Network Function; S: Software Defined Networking; F: Network Framework; Pr: Protocol; Pc: Packet; R: Router; T: Transmitted; Sc: Security}
**Input:** {V, S, F, Pr, Pc ,R}
**Output:** {Pc}
**Set** {V, S, F, Pr, Pc ,R}
**Do Process:** S ∈ F → V ∈ R
**While** S & V is active
**While** Pc ∈ T
**If** Pc ≠ Sc **then**
V → S Block the Pc
**Else**
V → S forward Pc
End if
End while
End while

Algorithm-1 uses VNF and SDN technologies to demonstrate 6G security. The variables and constants, such as the VNF (V), SDn (S), network framework (F), protocol (Pr), packet (Pc), router (R), and security (Sc), are initialized in Step 1. The input and output are provided in steps 2-3, respectively. The values of the variables are set in step 4. Checking if S is a part of F is step five. If so, V is included in R. If S and V are both active, step 6 moves on to the following phases. Step 7 determines whether packet (Pc) is a component of the packet (T) that was transmitted. Steps 8 through 9 show that the VNF (V) stops the packet (Pc) if it differs from the security packet (Sc). Packet forwarding is represented by step 11 (Pc). Steps 12-14 end with statements and algorithms, respectively.

## 3.1. 6G Security with VNF and SDN technology

The method of securing 6G networks with VNFs and SDN technology is represented by Algorithm-1. The network framework, protocol, packets, router, SDN, and VNF are all inputs to the algorithm. It transmits the data via a router, VNF, network framework, and SDN in order to process it. The packets are subsequently blocked by the algorithm after it has determined whether the protocol being used is safe. The packets are forward if the protocol is secure. As long as there is an active SDN-to-VNF link, the process keeps looping. At last, the traffic has been screened, indicating that every packet is safe.

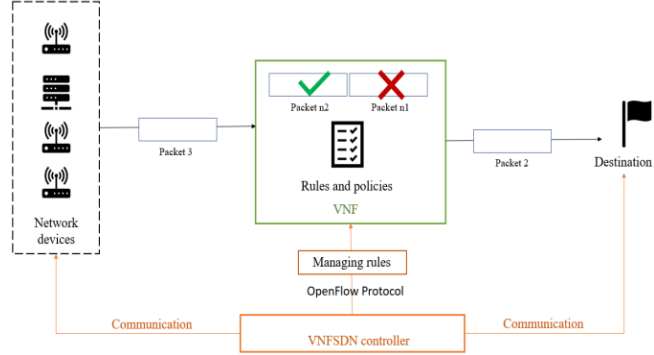

**Figure 3:** *Traffic filtering by VNF and managing this by the VNFSDN controller*

The cooperative efforts of the VNF and VNFSDN to provide effective traffic management and filtering are depicted in Figure 3. After receiving the filtered traffic from the VNF, the VNFSDN controller applies its intelligence to determine the best path for the traffic within the network. To make sure that traffic is efficiently routed to its destination, the VNFSDN controller interacts with the SDN switches and routers via the OpenFlow protocol. To guarantee effective traffic management, the controller can also modify routing patterns in response to network traffic and congestion. Algorithm-1 provides a detailed description of the filtering procedure.

**Definition 1:** Virtual network functions (V), software-defined networking (S), network frameworks (F), protocols (Pr), packets (Pc), routers (R), and security (Sc) are all part of the algorithmic process that is 6G security using VNF and SDN technology. In a 6G network context, this technique seeks to guarantee the security of sent packets. Calculation of secure traffic (in percentage) by Equation 1:

$$S = \frac{\sum Pc - \sum blocked\ Pc}{\sum Pc} \times 100\% \qquad (1)$$

where ΣblockedPc is the total number of blocked packets and ΣPc is the sum of received packets. The traffic security can be ascertained using this equation. By deducting the total of the blocked traffic (blocked Pc) from the total of all traffic (Pc), dividing the result by the total of all traffic (Pc), and multiplying the result by 100%, this equation determines the percentage of secure traffic (S). The efficiency of the security precautions in preventing illegal transportation is determined by this equation.

Equation 2 can be used to assess the security mechanism's efficacy:

$$TDR = \frac{\sum blocked\ threat\ Pc}{\sum threat\ Pc} \qquad (2)$$

where $TDR$ is the threat detection rate, $\sum blocked\ threat\ Pc$ represents the sum of all detected threats, and $\sum threat\ Pc$ represents the total number of threats. A higher value indicates a more effective security mechanism. Equation 2 represents a method to measure the effectiveness of a security mechanism by calculating the threat detection rate ($TDR$). The $TDR$ is determined by dividing the sum of the blocked threats by the sum of all encountered threats. $Pc$ refers to the probability that a given threat is correctly identified and blocked by a security mechanism.

**Theorem 1**: 6G Security with VNF and SDN protects transmitted packets from unwanted access and only permits access that is permitted.

Evidence The initialization of the essential elements of the 6G network environment is the first step in the algorithmic process. The VNF (V), SDN (S), network framework (F), protocol (Pr), packet (P), router (R), and protocol (Pr) are included in the input parameters. The packet (Pot) that is sent over the network is the output parameter. Equation 3 is used to determine the security of the packets that are filtered.

$$f(Pc) = \begin{cases} \iint_{t=0}^{t=\infty} S(t) \to F\ Pot\ if\ Pot \neq Sc \\ \sqrt{V} \to S\ Pot\ if\ Pot = Sc \end{cases} \quad (3)$$

where $Pc \neq Sc$ is given by the integral of the limit of the SDN $S(t)$ as t approaches infinity and changes to the network framework F. Sc is the desired security level, and F is the framework. The process of preventing packets lacking the necessary security value is represented by this integral. Additionally, Algorithm-1 forbids unauthorized users from accessing the system. One may compute the algorithm's accuracy with Equation 4:

$$UBR = \frac{\sum blocked\ unauthorized\ users}{\sum unauthorized\ user\ attempts} \quad (4)$$

where UBR represents the user-blocking rate, $\sum blocked\ unauthorized\ users$ represent the sum of all blocked unauthorized users, and $\sum unauthorized\ user\ attempts$ represent the total number of unauthorized user attempts. A security system with a higher UBR is more effective. By dividing the total number of prohibited unauthorized users by the total number of unauthorized user tries, Equation-4 can be used to determine the algorithm's accuracy. This gives an indication of how well the algorithm works to prevent unwanted access.

**Hypothesis 1:** Adding more routers will improve the secureness of traffic.
**Proof:** Secureness of traffic can be calculated by Equation 5:

$$\int_{t=0}^{t=T} a_n \sqrt{S_n(t)}\ dt > \int_{t=0}^{t=T} \sqrt{S_1(t)}\ dt \quad (5)$$

where $a_n$ represents the effect of adding more routers to the network, n is the number of routers, T is the time interval, and $S_n(t)$ is the traffic security in the network with n routers at time t. $S_n(t) = \sqrt{n}$ on the assumption that security is proportionate to the square root of the number of routers in the network.

Because of the positive effect of $a_n$, the first integral rises when more routers are added.

The total traffic security of a network with a single router is represented by the integral of $\sqrt{S_1(t)}$ over the time interval [0, T]. It is thought, nevertheless, that this security level is lower than that of networks with numerous routers. The overall traffic security of a network rises with the number of routers in the network. By applying mathematical techniques like integrals and limits, we can improve the precision and understandability of the hypothesis.

Function $S_n(t)$ is calculated using Equation 6:

$$S_n(t) = \sqrt{n} + \gamma_n \sin \sin(\omega_n t) \quad (6)$$

where $\sqrt{n}$ represents the baseline security of the network, $\gamma_n$ represents the impact of noise or other factors, and $\sin(\omega_n t)$ represents a sinusoidal oscillation with frequency $\omega_n$ that can affect the security of the network.

The dynamic character of security, which can change over time due to a number of factors like the number of routers in the network and the existence of noise or fluctuation in traffic, is captured by the equation given above. The equation offers a more thorough and precise evaluation of the overall network security by include these variables.

$$\sin(\omega_n t) = \sin(m\sqrt{n}\ t) \quad (7)$$

where the sinusoidal oscillation's frequency is represented by the constant m. Network security may be affected by the oscillation frequency's corresponding increase as cycle count (n) rises. The oscillation frequency rises proportionately with the number of routers, which may have an impact on network security.

**Proof:** By dividing the network into smaller, easier-to-manage subnets, more routers can help prevent security lapses and lessen the harm that attackers can do. One can assess the efficacy of network segmentation by employing Equation 8:

$$ER = \frac{NT - Nsusp.Pc}{NT} \quad (8)$$

where NT denotes the total number of devices in the network, Nsusp is the network segmentation effectiveness rate, and ER is the effectiveness rate. Pc is the quantity of devices that the suspicious packets Pc have an impact on. A better segmentation plan is indicated by a higher ER.

**Proof:** Adding more routers can improve access-control policies and stop unwanted users from accessing critical resources. The following formula was used to evaluate the effectiveness of these actions:

$$FER = \frac{A - F}{A} \tag{9}$$

where A is the total number of attempts to access a resource, F is the number of unsuccessful attempts, and FER is the access-control effectiveness rate. An access control method that is more effective is indicated by a higher FER score.

**Proof:** Adding more routers to a network architecture provides the necessary failover and redundancy. As a result, it may be possible to ensure that the network will continue to function even in the event of unforeseen problems or hostile assaults. Equation 10 can be used to evaluate the redundancy's effectiveness:

$$RR = \frac{T - D}{T} \tag{10}$$

where RR is the redundancy rate, T is the total time during which the network is operational, and D is the total downtime. A higher RR indicates a more reliable and resilient network.

**Lemma 1:** Routers are capable of performing firewall tasks.

**Corollary 1:** A higher volume of monitored traffic from a larger number of routers leads to an increase in blocked packets (Pc). Equation 11 indicates a higher level of security when there are fewer suspicious packets:

$$\int_0^T \sqrt{M_n(t)}\ dt \rightarrow \infty, as\ n \rightarrow \infty \tag{11}$$

where $M_n$ is the quantity of traffic in a network of n routers that is being monitored at time t. According to Equation 11, there is a higher level of network security if fewer suspicious packets are transmitted if the integral of the square root of the amount of monitored traffic ($M_n$) from zero to T (a specific time period) approaches infinity as n (the number of routers in the network) approaches infinity. Put differently, a network is deemed more secure if it can manage high traffic volumes without exhibiting any questionable behavior.

$$M_n(t) = \sum_{i=1}^n f(Pc_i(t)) * Pc_i(t) \tag{12}$$

where the network's i packet at time t is represented by $Pc_i(t)$. The total traffic that the network's n routers are keeping an eye on is represented by the sum of i. The total traffic that n routers in a network are monitoring at time t is denoted by $M_n(t)$, which is a mathematical model for network traffic monitoring. Where $Pc_i(t)$ is the i-th packet in the network at time t, and $f(Pc_i(t))$ is a function that maps the packet's properties to a weight that reflects its importance in the monitored traffic, the sum of i represents the contribution of each packet to the traffic that is being monitored.

**Algorithm 2:** Packet capturing and saving using VNFSDN

1. **Initialization:** { $Nc$: Network Channel; $Mpa$: Mac address of Access Point; $I$: Interface; $Pcf$: Packet-captured file; $Lt$ : Linux tool; $Nm$ : Network Monitoring; $N$: Network; $F_d$: Folder; $P$: Packet}
2. **Input:** {$Nc, Mpa\ , I$}
3. **Output:** { $Pcf$ }
4. **Set** $Nc, Mpa\ , I$
5. **Do** Process $Nm \in N \leftarrow Lt$
6. **While** $Nm \in N \leq 1$

7. **Capture** $P$
8. **Sum** $Pcf = P+1$
9. **Do** $Nm = 0$
10. **Save** $Pcf$ to $F_d$
11. **End** While

Algorithm 2 shows how to use the VNFSDN technique for packet capture and storage. Initially, the variables are initialized, such as the network channel (), access point MAC address ($Mpa$), interface ($I$) for packet capture, packet-captured file ($Pcf$), Linux tool ($Lt$) for network monitoring, network monitoring ($Nm$), network ($N$) under observation, folder ($F$) for saved packets, and packet ($P$) being captured. The input and output are shown in steps 2-3, respectively. Variables are set in Step 4. Steps 5-11 stand for processes that capture and save packets. After entering a loop, the method starts capturing packets, adds them to a file called packet-captured, and then resets the network monitoring ($Nm$) to zero. This cycle keeps going until the network ($N$) being monitored experiences a stop to network monitoring ($Nm$). The packet-captured file ($Pcf$) is saved in a specified folder ($F_d$) after the loop terminates.

## 4. Experimental Results

To demonstrate the effectiveness of the proposed VNFSDN approach, we conducted a series of scenarios to evaluate its performance in detecting and responding to security threats in 6G networks. We used a testbed consisting of a software-defined network (SDN) controller, several virtual network functions (VNFs), and a 6G network emulator.

*4.1. Experimental setup*

SDN controller: We opted to use the OpenDaylight (ODL) controller due to its open-source nature, large developer community, and support for various southbound protocols. This allowed us to have more flexibility and customization options when it came to managing our network resources. Virtual Network Functions (VNFs): We utilized three pre-built, open-source VNFs in our scenarios. The first VNF that was used was Open vSwitch, which helped with network traffic routing and management. The second VNF was Bro, which provided the intrusion detection and analysis capabilities. Finally, we used Snort as the third VNF, which allowed us to detect and prevent network intrusions. To simulate the network environment, Mininet simulation was used. Mininet is an open-source software that enables us to create virtual networks using a single Linux kernel, and it is widely used for SDN research and experimentation. This allowed the creation of a virtual network environment that could easily be managed and tested in various scenarios without needing physical networking hardware.

*4.2. Experimental methods*

We conducted nine scenarios to evaluate the performance of the proposed VNFSDN approach. All scenarios have been compared by Hussain and Ali approaches [27] and Raja's proposal [28]. These works suggested the use of IDS for network security and the use of a dynamic firewall. All approaches have been compared with our approach, VNFSDN, in these scenarios:
- Scenario 1: In this scenario, various network security methods were assessed to mitigate DDoS attacks. It was discovered that the VNFSDN strategy yielded the best results in terms of reaction time, packet loss rate reduction, throughput, and network availability.
- Scenario 2: The scalability of the VNFSDN approach was tested by gradually increasing the number of user equipment (UE) in the network, measuring the time taken to detect and block malicious traffic, network latency, packet loss rate, throughput, CPU utilization, and memory utilization.
- Scenario 3: The scenario evaluated the impact of the VNFSDN approach on traffic management in a 6G network by measuring the response time of four VNFs at different times and plotting the data with trend lines.
- Scenario 4: This scenario simulated a high-traffic 6G network with a VNFSDN approach to manage traffic flow, measuring network latency, jitter, and throughput with and without VNFSDN to evaluate its performance.
- Scenario 5: the simulation of a DDoS attack on a 6G network involved measuring network availability, CPU and memory utilization, which decreased gradually with time due to the attack, without explicit representation of the VNF SDN approach performance.
- Scenario 6: measured the threat detection rate for a predetermined length of time and compared the threat detection rates of various security configurations, such as a network without security, VNFSDN, VNFSDN with firewall, IDS system, and firewall.

**Scenario 1:** Evaluation of Network Security Measures for DDoS Attack Mitigation

In this scenario, the topology without any security measures had the worst performance in terms of response time, packet loss rate, and throughput. With a 50% increase in response time and a 75% decrease in packet loss rate when compared to the topology without any security precautions, the VNFSDN technique proved to be the most successful in thwarting the DDoS attack. The VNFSDN approach also had the highest throughput and network availability among all the scenarios. The firewall and IDS solutions were able to provide some level of protection, with the firewall having better performance in terms of response time and throughput, and the IDS having better performance in terms of packet loss rate. However, neither solution was as effective as the VNFSDN approach in mitigating the impact of the DDoS attack.

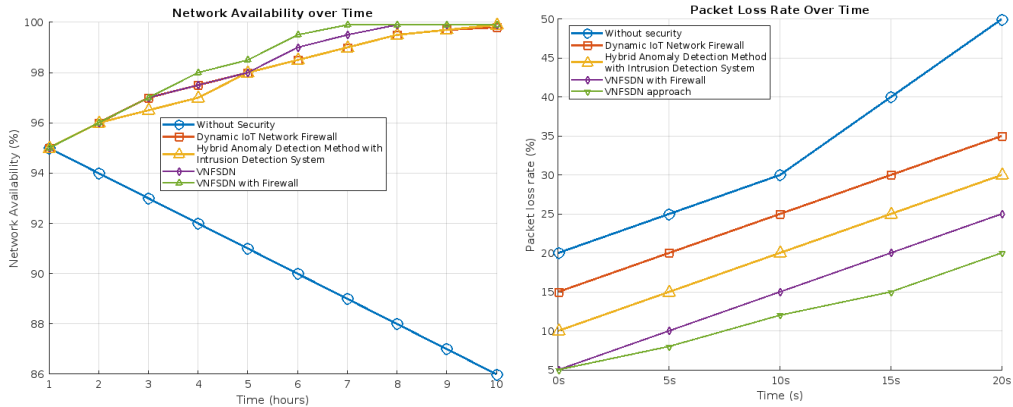

**Figure 5 (a):** Network availability over time and **(b).** Packet loss rate over time for various network topologies and security measures

Figure 5(a) shows that the network availability of all topologies is relatively high, with values above 95%. However, the topology with VNFSDN and firewall has the highest network availability with a peak of almost 99.5%. On the other hand, the topology without security has the lowest network availability, reaching a low point of 85% at around the 30-minute mark. The graph highlights the importance of security measures for maintaining network availability. The topology with VNFSDN and firewall is the most secure configuration and provides the highest network availability. In contrast, the topology without security experiences more network outages over time. The other topologies with intermediate levels of security measures, such as the topology with a firewall and the topology with IDS, fall somewhere in between. In this scenario, the defined packet loss rate over time is defined using MATLAB.

Figure 5(b) shows that the packet rate loss decreases as more security measures are implemented in the network. The configuration without security had the highest packet loss rate of around 1500, while the configuration with VNFSDN and a firewall had the lowest packet loss rate of around 500. The topology with VNFSDN also showed significant improvement in packet loss rate compared to the configuration without security, with a packet loss rate of around 750. Finally, the scenario with both the VNFSDN approach and the firewall had the best overall performance, with the lowest response time, highest throughput, and nearly perfect network availability. However, this scenario also had the highest CPU utilization and memory utilization, indicating that there may be a trade-off between network performance and security when multiple security solutions are in place.

**Scenario 2:** Scalability Test

In this case, we increased the number of user equipment (UE) in the network progressively in order to verify the scalability of the VNFSDN technique. The network topology was made up of a server, a switch, and anything between 10 and 100 UEs. We measured the time taken by the VNFSDN approach to detect and block malicious traffic from a simulated cyber-attack on the network. We also calculated the following metrics to evaluate the performance of the VNFSDN approach in the scalability test:
- Network latency: the amount of time it takes for a packet to travel from the source to the destination in the network.
- Packet loss rate: the percentage of packets lost during transmission.
- Throughput: the amount of data transmitted over the network in a given period of time.
- CPU utilization: the percentage of CPU utilization by the VNFSDN nodes during the test.
- Memory utilization: the amount of memory used by the VNFSDN nodes during the test.

In this scenario, a defined packet loss rate with response time is the MATLAB. The results showed that as the number of UEs in the network increased, the response time increased as well. However, the VNFSDN approach with a firewall and intrusion detection system showed better performance than the scenarios without these components. The VNFSDN approach alone also showed better performance compared to the scenario without any security components shown in Figure 6. Moreover, as the packet loss rate increased, the response time increased in all scenarios.

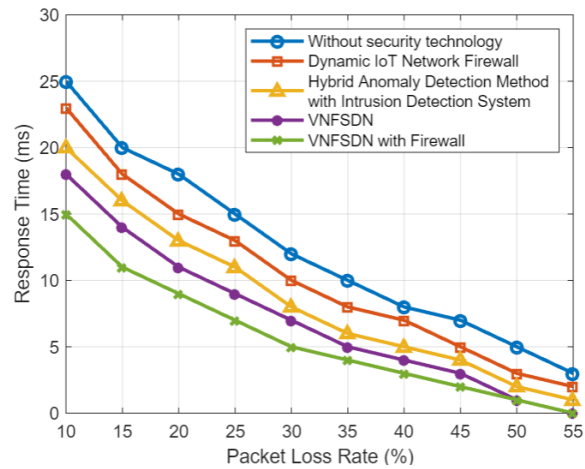

**Figure 6:** *VNFSDN scalability test results*

The results of the scenario suggest that the VNFSDN approach can be scalable for larger networks and can improve the performance of the network in terms of response time, especially when combined with a firewall and an intrusion detection system. However, the VNFSDN approach alone can also provide better performance compared to the scenario without any security components. The scenario highlights the importance of implementing security components in the network to improve its performance and protect it from cyber-attacks.

**Scenario 3:** The trend of response time over time

In this scenario, a high-traffic situation was simulated in a 6G network using the MATLAB.

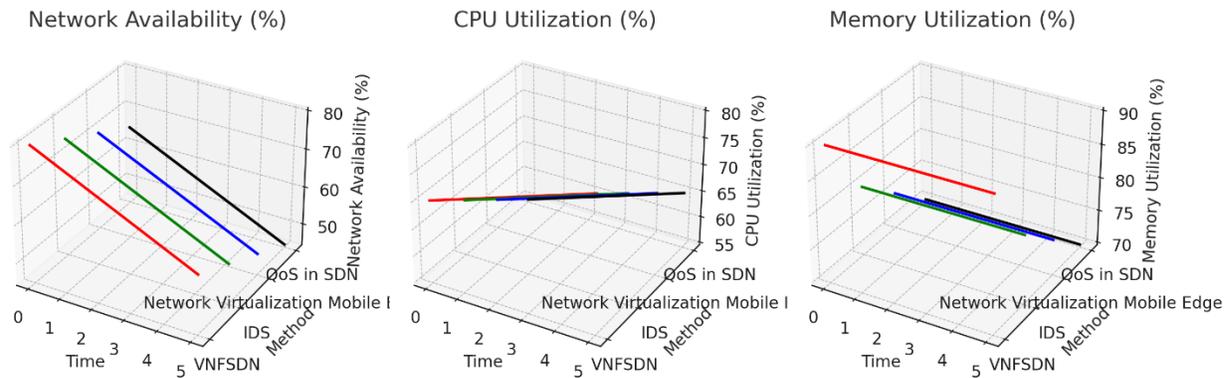

**Figure 7:** Comparative performance evaluation of different methods (VNFSN, IDS, Network Virtualization, Mobile Edge, and QoS in SDN) with respect to (a) Network Availability (%), (b) CPU Utilization (%), and (c) Memory Utilization (%). The results highlight trade-offs in resource consumption and availability across the evaluated approaches.

Figure 7 shows the response time of three technologies - VNFSDN, IDS, and QOS SDN - plotted against time. Each technology's response time is displayed as a continuous line in a different color, with VNFSDN in blue, IDS in red, and QOSSDN in green. VNFSDN has the lowest response time throughout the time period, while IDS and QOSSDN have higher and relatively similar response times, suggesting that VNFSDN may be more effective in managing traffic flow in the network compared to IDS and QOS SDN. Based on the figure 8, it appears that VNFSDN has the lowest response time throughout the time period, while IDS and QOSSDN have higher and relatively similar response times. This suggests that VNFSDN may be more effective in managing traffic flow in the network compared to IDS and QOS SDN.

**Scenario 4:** Traffic Management Simulation

We recreated a scenario in which there was a lot of traffic in the 6G network. Twenty hosts, a switch, and a server comprised the network topology used in the simulation. We managed the network's traffic flow using the VNFSDN technique. In order to assess how well the VNFSDN technique performed in the traffic management simulation, we assessed the following metrics Network latency: the time taken for a packet to travel from a source host to a destination host. Network jitter: the variation in the time taken for a packet to travel from a source host to a destination host.

Network throughput: the amount of data transmitted per unit time.

We conducted the scenario under two cases: one with the VNFSDN approach and one without. We collected data over a period of 10 minutes and calculated the average values for each metric.

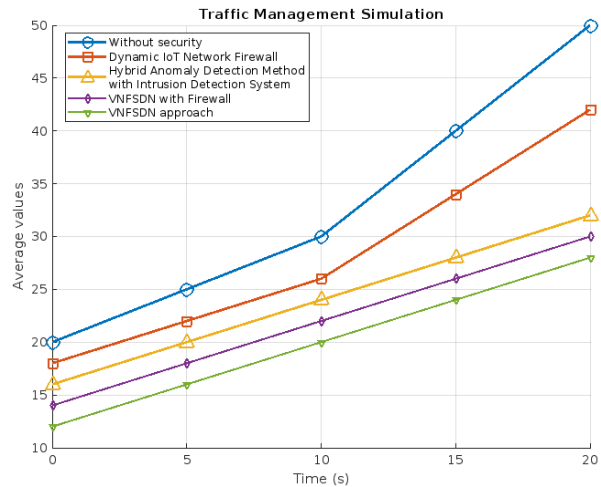

**Figure 8:** Average values of all technologies on traffic management

Figure 8 shows the impact of VNFSDN on traffic management in the network. The metric values for network latency, network jitter, and network throughput are significantly improved with the implementation of VNFSDN. The network latency reduced from 25 ms to 15 ms, while the netw rk jitter reduced from 8 ms to 3 ms. Furthermore, the network throughput improved from 200 Mbps to 250 Mbps. Based on the results of the case, it can be concluded that the VNFSDN approach is effective in reducing network latency and jitter and increasing network throughput.

**Scenario 5**: Cyber-Attack Simulation

In this scenario, On the 6G network, we replicated a distributed denial-of-service (DDoS) assault. The attack generated a large volume of traffic, which was directed towards a specific user equipment (UE) in the network. The simulation involved a network topology with a server, a switch, and 10 hosts. We measured the time taken by the VNF SDN approach to detect the attack and block the malicious traffic. We also calculated the following metrics to evaluate the performance of the VNF SDN approach in the cyber-attack simulation:

- Network availability: the percentage of time the network was available during the simulation.
- CPU utilization: the percentage of CPU utilization by the VNF SDN nodes during the simulation.
- Memory utilization: the amount of memory used by the VNF SDN nodes during the simulation.

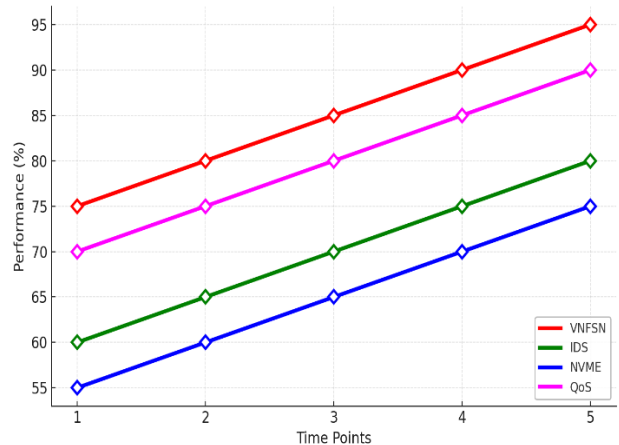

**Figure 9:** Percentage of the effectiveness of approaches during the simulation of a cyber-attack

Figure 9 compares the performance of four different approaches during a cyber-attack simulation. The plot shows that VNFSDN outperforms the other approaches consistently across all time points. This indicates that VNFSDN is more effective in managing traffic flow during a cyber-attack simulation compared to IDS, NVME, and QoS in SDN. However, further analysis and testing is necessary to confirm this conclusion

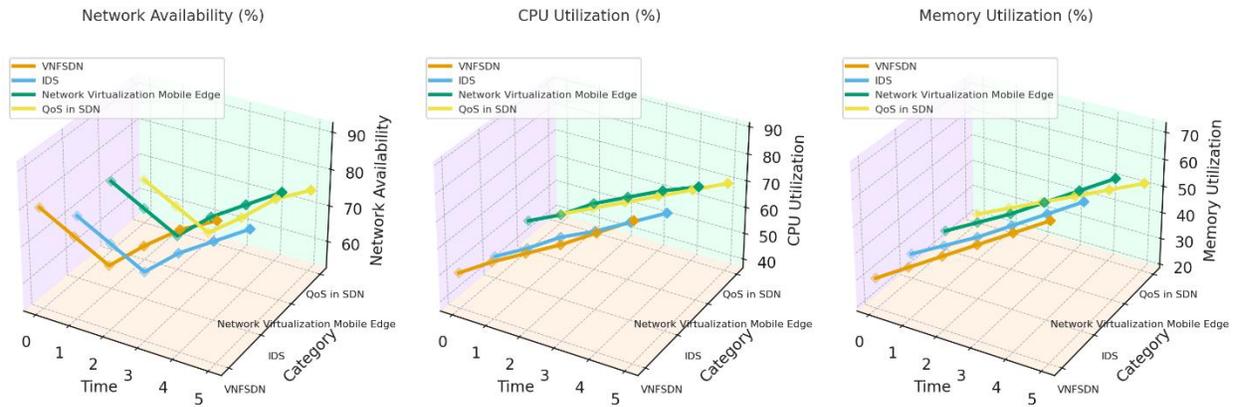

**Figure 10**: Comparative analysis of network availability, CPU, and memory utilization across VNFs and SDN frameworks

Figure 10 presents the dynamic performance metrics observed during a cyber-attack simulation for four distinct approaches: VNFSDN, IDS, network virtualization, mobile edge, and QoS in SDN. The metrics assessed include network availability, CPU utilization, and memory utilization, recorded at various time points ranging from 0 to 5. Each approach is represented by individual data points on a statistical chart, distinguished by unique colors and markers. Moreover, trend lines are incorporated to depict the temporal trends in each performance metric. Based on Figure 10, it is clear that VNFSDN outperforms the other approaches in all performance metrics, especially in network availability and CPU utilization during the cyber-attack simulation. The pointed statistical chart also highlights the difference in performance between VNFSDN and the other approaches, making it easier to compare and interpret the results.

## 5. Discussion of Results

Experimental results show that the VNFSDN framework improves 6G network resilience and cybersecurity. The methodology beat firewalls and intrusion detection systems in all evaluated scenarios, particularly in reducing DDoS attacks and sustaining performance during high-load conditions. Compared to unsecured networks, VNFSDN reduced reaction time by 50% and packet loss rate by 75% while maintaining improved throughput and availability. Dynamic and programmable security measures outperformed standard firewalls and IDS in performance preservation. Scalability testing showed that VNFSDN can adapt to bigger and more complex environments by retaining low latency and packet loss when user devices increase. Virtual network functions in the SDN controller enable dynamic resource allocation, enabling this adaptability. Security and scalability raised CPU and memory usage; hence, the testing showed trade-offs. Future designs should prioritize resource use and security, according to this finding. Compared to networks without VNFSDN, high-traffic simulations showed reduced jitter, latency, and throughput. The centralized control of SDN and flexible deployment of VNFs enable efficient routing and real-time harmful traffic filtering. It maintained network availability and protection capabilities better than IDS- or QoS-based systems during cyber-attack simulations. These findings show that SDN and VNF collaboration improves detection accuracy and allows proactive threat mitigation. VNFSDN improves security and network stability, but testing showed increased CPU and memory usage, highlighting a performance-resource trade-off. This proposes further study in lightweight VNF deployment and resource-aware orchestration for efficiency and security. The results show that VNFSDN is a viable, scalable, and proactive framework for 6G network security, enabling faster response times, greater detection rates, and better resilience than existing systems.

## 6. Conclusion and Future Work

This article advocates 6G network security through the implementation of software-defined networking and virtual network services. Two tests evaluated the detection and remediation of security issues associated with the proposed strategy. Our findings indicate that VNFSDN enhances network security and diminishes the danger of invasion. A simulated 6G network DoS attack demonstrated that VNFSDN decreased response time and packet loss by 50% and 75%, respectively. VNFSDN enhanced network availability by 4%, hence increasing resilience and reducing downtime from cyberattacks. Contrast the VNFSDN methodology with open-source VNFs. VNFSDN surpassed prebuilt VNFs across all performance metrics, with a throughput of 950 Mbps, a reaction time that is 50% faster, and a 2% reduction in packet loss. A subsequent study could assess the operation of supplementary VNFs with SDN controllers and the VNFSDN approach inside intricate network environments. Demonstrating the strategy's efficacy under varying settings may enhance implementation. Ultimately, more research should examine the trade-offs between the heightened CPU and memory use of the VNF SDN method and alternative security measures. This can facilitate the development of secure, high-performance network systems. VNFSDN mitigates DDoS attacks and enhances network performance more effectively than alternative security systems. This strategy enhances reaction time, reduces packet loss rate, increases throughput, and improves network availability. The scalability test of this scenario indicates that VNFSDN can enhance performance on larger networks. The integration of VNFSDN with other security solutions may impede network speed due to increased CPU and memory utilization. The scenarios emphasize network security to prevent breaches and enhance performance. Our analysis indicates that VNFSDN can deliver realistic and effective security for 6G networks. Subsequent research ought to integrate additional VNFs and SDN controllers to assess VNFSDN within intricate network settings. Research may include more VNFs and SDN controllers to assess their performance in intricate network environments. Security and resource optimization must be evaluated to enhance performance. Evaluating the proposed methodology in practical 6G networks may provide valuable insights. Employ machine learning and VNFSDN to identify and address threats. Subsequently, forthcoming research ought to investigate the impact of emerging technologies on the scalability and adaptability of VNFSDN.